\documentclass[a4paper,11pt]{article}
\pdfoutput=1 
\usepackage{jheppub} 
\usepackage{epsfig}
\usepackage{caption}
\usepackage{yfonts}
\usepackage{subcaption}
\usepackage{color,soul,amsmath}
\usepackage{slashed}
\usepackage{float}
\usepackage{amsfonts}
\usepackage{url}
\usepackage{slashed}
\usepackage{accents}
\usepackage{bbm,multicol}
\usepackage{lipsum}
\usepackage{mathtools}
\usepackage{physics}

\usepackage[normalem]{ulem}

\hypersetup{
  bookmarks=true,         
  unicode=false,          
 pdftoolbar=true,        
 pdfmenubar=true,        
 pdffitwindow=true,     
 pdfstartview={FitH},    
 pdfsubject={Dark Matter},   
 pdfnewwindow=true,      
 pdfcreator={RevTeX},
 colorlinks=true,       
 linkcolor=red,          
 citecolor=blue,        
 filecolor=black,      
 urlcolor=blue,           
  }

\begin{document}
\begin{center}
{\Large\bf Prospects of measuring the atmospheric muon neutrino and anti-neutrino flux ratio with the ATLAS detector}\\
\end{center}

\begin{center}
{\bf Deep Ghosh}\footnote{E-mail: matrideb1@gmail.com},
{\bf Satyanarayan Mukhopadhyay} \footnote{E-mail: tpsnm@iacs.res.in}
{\bf and Biswarup Mukhopadhyaya}\footnote{E-mail: biswarup@iiserkol.ac.in}\\ \bigskip
{\em $^{1,3}$Department of Physical Sciences, Indian Institute of Science Education and Research (IISER) Kolkata, Campus Road, Mohanpur, West Bengal 741246.\\ \bigskip
$^2$School of Physical Sciences, Indian Association for the Cultivation of Science (IACS), 2A and 2B Raja S.C. Mullick Road, Kolkata 700 032.}
\\[20mm] 
\end{center}

\begin{center}
\underline{\Large{\textbf{Abstract}}}
\end{center}

There is a significant uncertainty in the prediction of atmospheric muon neutrino and anti-neutrino flux ratio using different flux models, especially at higher energies. We study the prospects of experimentally measuring this flux ratio as a function of energy with the ATLAS detector at the LHC. To this end, we compute the contained-vertex and external upward going charged current event rates induced by atmospheric muon (anti-)neutrinos through deep inelastic scattering at the 4 kiloton hadron calorimeter (HCAL) component of ATLAS. We illustrate the event selection criteria necessary to eliminate the cosmic ray muon background for the above event classes. While the contained vertex events have a striking topology with a muon being created inside the HCAL and then travelling to the muon chamber possibly through the tracker, for muons with energy larger than 3 GeV, nearly $10$ times more events are obtained for the external upward going muons created in the rock column below the detector. Our estimates show that the energy dependence of the ratio of negative and positively charged muons induced by atmospheric muon neutrino and anti-neutrino fluxes can be measured by ATLAS upto a muon energy of 100 GeV, with 1000-live days of neutrino physics exposure over a period of several years, considering only the period with the LHC beams not in circulation, but the detector and magnetic fields of ATLAS in operation. With this exposure, we expect to obtain $60~\mu^-$ and $30~\mu^+$ contained vertex events, and $599~\mu^-$ and $292~\mu^+$ external upward-going events, after imposing the necessary selection criteria. For the latter class of events, this corresponds to an expected ratio of negative to positive charged muon events  averaged over all energies,  $R_{\mu^-/ \mu^+}=2.05^{+0.15}_{-0.14}$, at $68\%$ C.L. The CMS detector at the LHC can also be used with comparable reach for studying the external upward-going rock muon events. 

\vskip 1 true cm

\newpage

\section{Introduction}
\label{sec:intro}
In interpreting the results of atmospheric neutrino experiments, and determining the neutrino mass and mixing angles using them, an important ingredient is the knowledge of the atmospheric neutrino fluxes. These fluxes are computed using advanced numerical flux models, which utilize the primary cosmic ray spectrum, the expected yield of neutrinos from cosmic ray interactions with the atmosphere, as well as data on cosmic ray muons at sea level to determine certain parameters of the models~\cite{Gaisser_Book, Volkova:1980sw}. 

Atmospheric neutrinos are produced in the interaction of primary cosmic rays (around 90\% protons, 9\% alpha particles and a small percentage of heavier nuclei) with the air molecules in the atmosphere, such as nitrogen and oxygen~\cite{Gaisser_Book}. Depending on the energy of the primary proton, different mesons can be produced in these interactions, such as, $p+{\rm Nitrogen} \rightarrow \pi^\pm, \pi^0, K^\pm, K_L, K_S, ...$ These mesons subsequently decay to generate a flux of atmospheric muons and neutrinos of different flavours. At low energies, the most copiously produced mesons are charged and neutral pions, where the charged pions decay before reaching the sea-level (a 10 GeV pion travels $\sim 0.5$ km before decaying), with $\pi^+ \rightarrow \mu^+ \nu_\mu$ and $\pi^- \rightarrow \mu^- \overline{\nu}_\mu$. If the muon energy is less than $\sim 2.5$ GeV, it also decays within the $\sim 15$ km length of the atmosphere by the standard decay chains: $\mu^+ \rightarrow e^+ \nu_e \overline{\nu}_\mu$ and $\mu^- \rightarrow e^- \overline{\nu}_e  \nu_\mu$. 

Therefore, for energies of the order of a GeV or so, we expect both $\pi^+$ and $\pi^-$ decays to produce equal numbers of $\nu_\mu$ and $\overline{\nu}_\mu$, as long as the muons decay on their way to the sea-level, with the ratio:
\begin{equation}
R_{\nu_\mu/ \overline{\nu}_\mu} = \frac{N_{\nu_\mu}}{N_{\overline{\nu}_\mu}} \simeq \mathcal{O} (1), {~~~\rm upto} ~E_\nu \sim 1 {~\rm GeV}.
\end{equation}
For higher energies, the muons do not decay before reaching sea-level, and the ratio starts to increase, as more $\pi^+$ are produced compared to $\pi^-$ in proton interactions with air molecules, given the respective quark contents. 
At even higher energies, $K-$mesons are also produced in increasingly larger numbers, with more $K^+$ than $K^-$, with the following dominant decay modes: $K^+ \rightarrow \mu^+ \nu_\mu (64\%)$, $\pi^+ \pi^0 (21 \%)$, $\pi^0 \mu^+ \nu_\mu (3 \%)$, $\pi^+\pi^+\pi^- (6\%)$ and $\pi^+ \pi^0\pi^0 (2\%)$. Combining these inputs, we see that the ratio $R_{\nu_\mu/ \overline{\nu}_\mu}$ starts from 1 at low energies, and then grows at higher energies. Its value, averaged over all zenith angles is around 1.2 at 10 GeV neutrino energy, 1.4 at 100 GeV, 1.5 at 1 TeV, etc, using the flux model of Honda et al~\cite{Honda:2015fha, Honda:2011nf}. 

However, as we can see from Refs.~\cite{Honda:2015fha, Honda:2011nf,SK_Full}, significant uncertainties in the muon 
(anti-)neutrino flux ratio $R_{\nu_\mu/ \overline{\nu}_\mu}$ exists, when we compare the predictions of different flux models, with differences of $25\%$ or higher at neutrino energies of order 100 GeV~\cite{Battistoni:2003ju, Honda:2004yz, Barr:2004br}. {\em This significant difference between neutrino and anti-neutrino fluxes may not only impact atmospheric neutrino physics analyses, they are sensitive to the flux model parameters that may feed into the total neutrino flux as well. The goal of this paper is to discuss experimental measurements of this ratio, which, when included in the fits, may be helpful in reducing the uncertainties of the atmospheric flux models.} In this connection, the charge ratio of cosmic ray muons has been measured earlier, see for example, Refs.\cite{Super-Kamiokande:2024rwz,Yanez:2019bnw}. Most neutrino detectors, such as Super-Kamiokande (SK) and IceCube, primarily measure the total neutrino-antineutrino flux \cite{Super-Kamiokande:2015qek,IceCubeCollaboration:2023wtb}. However, in SK, antineutrinos can be identified via neutron-tagging \cite{Beacom:2003nk}, although this method works for antineutrinos of energy less than around $30$ GeV, and in addition the efficiency is rather low \cite{Super-Kamiokande:2022hxq,Super-Kamiokande:2023ahc}. In this paper, \textit{we propose to extend these previous studies by tagging both neutrinos and antineutrinos at ATLAS with a larger efficiency}. 

As is well-known, although neutrinos are the most abundant cosmic rays at the sea level, their detection is hard due to small neutrino-nucleon scattering cross-sections. For example, the cross-section for producing a charged lepton (averaged over neutrino and anti-neutrino) in a broad energy range of 1 GeV to 3000 GeV is approximately $0.5 \times 10^{-38} {~\rm cm}^2 \times E_\nu (\rm GeV)$. On the other hand,  the neutrino flux around 1 GeV energy, summed over all directions is around $1 {~\rm cm}^{-2} {\rm s}^{-1}$. Thus, for 1 GeV atmospheric neutrinos one expects around $100$ neutrino interactions in a detector of fiducial mass 1 kiloton with one-year of exposure. Hence, to study the charged current interactions of neutrinos, we need a detector of at least few kilotons fiducial mass, running for few hundred live days. In addition, to distinguish neutrinos from anti-neutrinos using charged current processes, we need a detector with an ability to distinguish a charged lepton from an anti-lepton, possibly with a magnetic field. 

The MINOS experiment had all of the above features~\cite{MINOS:2008hdf, Gaisser:2002zv}. Although the MINOS experiment was operated primarily using a neutrino beam, it carried out a study of atmospheric neutrinos when the beam was not in circulation.  The MINOS far detector had a mass of 5.4 kton, but only around 4 kton fiducial mass was available for atmospheric neutrino studies. It had a magnetic field of 1.3 T in the far detector, making neutrino vs anti-neutrino studies feasible. Measurements of atmospheric neutrino and antineutrino interactions in the MINOS Far Detector were made, based on 2553 live-days of data~\cite{MINOS_Atmospheric}. A total of 2072 candidate events were observed. These were classified into 905 contained-vertex muons and 466 neutrino-induced rock-muons, both produced by charged-current interactions. For contained vertex events, they reported a ratio of muon neutrino to anti-neutrino of about 2.2 (with a 10\% statistical error) and for neutrino-induced rock muons a ratio of 1.6 (with a 15\% statistical error)~\cite{MINOS_Atmospheric}. Energy dependence of this ratio was not reported by MINOS. 

\begin{figure}[htb!]
\centering
\includegraphics[scale=0.3]{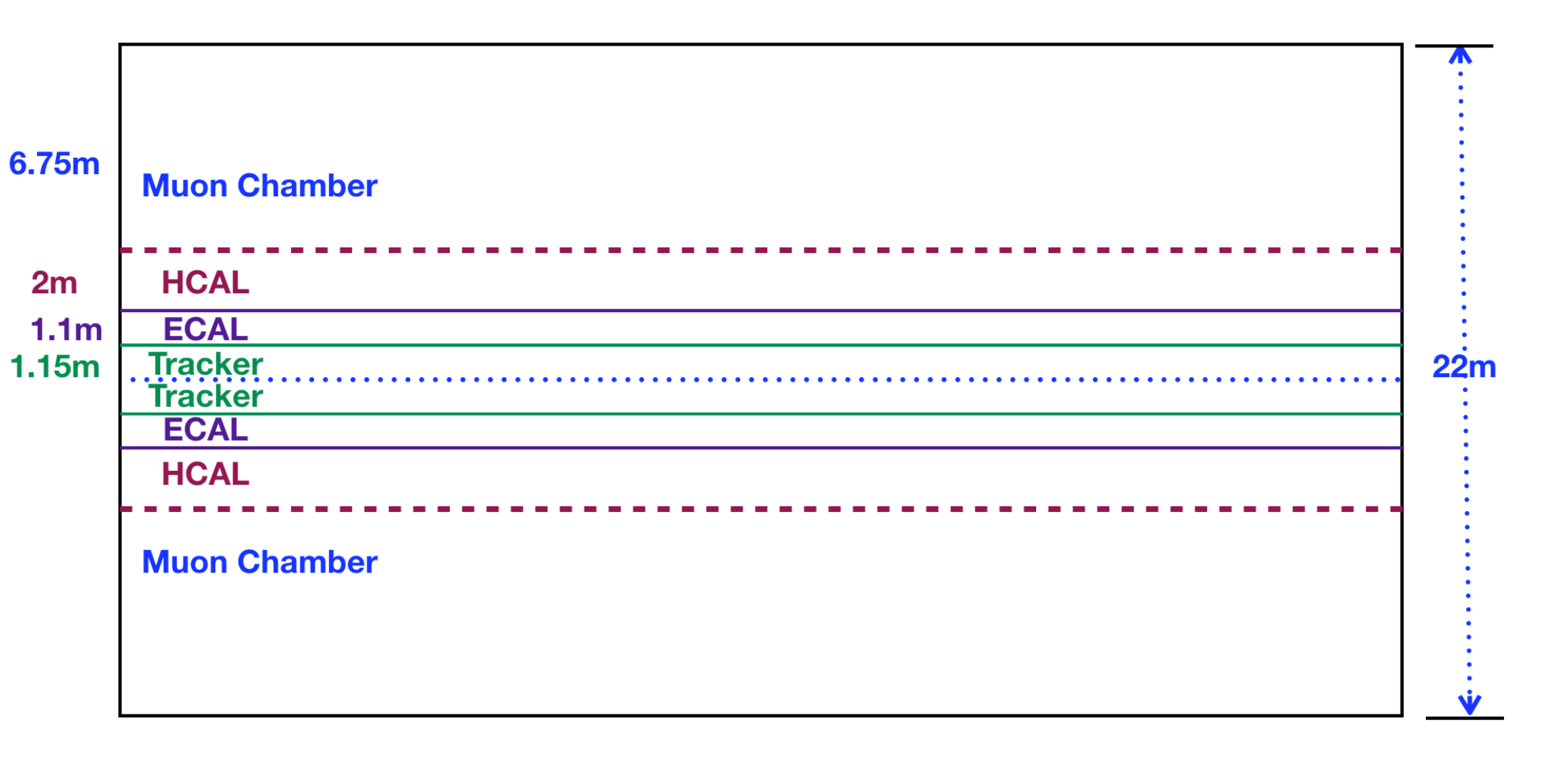}
\caption{{\em A schematic outline of the cross-sectional view of the ATLAS detector at the LHC following Ref.~\cite{ATLAS:2008xda}, where the width of the tracker, ECAL, HCAL and muon chambers are indicated, on both sides of the central beam axis. For the current study, it is important to note that the muon chamber is of $\sim 7$ m width on both sides, and hence a muon produced at the HCAL needs at least around 23 ns time to reach the end of the muon chamber.}}
\label{fig:ATLAS}
\end{figure}
{\em In this paper, we propose to experimentally measure the muon neutrino and anti-neutrino flux ratio using the ATLAS detector at the CERN Large Hadron Collider (LHC), which is the largest collider detector ever built.} While the overall weight of ATLAS is around 7000 tons, the weight of the ATLAS hadron calorimeter (HCAL), which is mostly made of steel with plastic scintillators, is around 4000 tons~\cite{ATLAS:2008xda}. In this study, we shall be exploring the ATLAS hadron calorimeter as the primary fiducial mass for atmospheric neutrino induced contained vertex charged current events. The magnet system at ATLAS includes a 2 Tesla solenoid for the inner detector,  a 0.5 Tesla toroid for the barrel and a 1 Tesla toroid for the muon end-caps. The width of ATLAS is around 22 meters, out of which the width of the tracker, electromagnetic calorimeter (ECAL), HCAL, and muon chambers being 1.15 m, 1.1 m, 2 m and 6.75 m, respectively, on both sides of the beam line~\cite{ATLAS:2008xda}. Since these length scales quantifying the dimensions of different detector components are relevant to our study, we present a schematic outline of the cross-sectional view of the ATLAS detector in Fig.~\ref{fig:ATLAS}.

Given the fact that the ATLAS HCAL has the necessary minimum fiducial mass to carry out atmospheric neutrino measurements, the next question is the exposure time available. The ideal configuration for neutrino studies will be with the LHC beams OFF, and the detector and magnetic field ON. A considerable amount of cosmic ray muon data were already collected in this configuration at the beginning of Run-1 and Run-2 -- for example, for ATLAS, during 2008-2010~\cite{ATLAS:2011zty}. The LHC beams are not in circulation during the winter months, while the detector and magnetic fields are ON during some days of this period (for cosmic ray studies and detector alignment and other checks)~\cite{Kopp:2007ai,S_Banerjee}. Exactly how many days of data have been recorded in this configuration every year can only be confirmed by the ATLAS and CMS cosmic ray groups. Furthermore, for the year 2024, the approved duration for physics run is around 172.5 days (including all of p-p and Pb-Pb runs). Excluding the winter shutdown period, there will be 44 days for beam commissioning, scrubbing and intensity ramp up, 26 days for machine development sessions and 10 days for technical stops and recovery~\cite{Steerenberg}. It is conceivable that out of these ~80 days in a year, a fraction might be used for neutrino studies.  Finally, during the beam intensity ramp up phase, the number of pile-ups may be low, and with a primary vertex veto (no collisions), we may also look for atmospheric neutrino events~\cite{Kopp:2007ai, Steerenberg}. In what follows, we shall assume a total of 1000 live-days for neutrino physics studies, gathered over several years from 2008 to the lifetime of the LHC experiments.

The idea of carrying out neutrino physics studies with a large and finely instrumented detector such as ATLAS was first proposed by F.~Vannucci privately to different authors~\cite{Kopp:2007ai,Vannucci,Petcov}, and finds its first passing mention in a study of magnetized iron detectors by Petcov and Schwetz (2006)~\cite{Petcov}, although they did not perform any analysis. Kopp and Lindner (2007)~\cite{Kopp:2007ai} carried out the first analysis of using ATLAS for atmospheric neutrino oscillation studies with the aim of more precise neutrino mass and mixing estimations. Their study used a minimum reconstructed neutrino energy of 1.5 GeV, which does not seem to be feasible with the ATLAS detector. As we shall subsequently discuss, a muon of minimum energy 3 GeV is necessary to obtain a signature in the muon chamber, after it successfully crosses the HCAL without losing all of its energy. Recently, Wen et al (2024)~\cite{Wen:2023ijf} looked into the prospects of observing high-energy supernova neutrinos from a particular direction in the sky, with a limited yield of $10-100$ through-going events at ATLAS. As we shall discuss in detail later, the main background of neutrino-induced muon events comes from cosmic-ray muons. The study of cosmic-ray muons in collider detectors was performed in LEP experiments \cite{Ridky:2005mx,L3:2004sed}, and recently by the ALICE collaboration \cite{ALICE:2015wfa} with multi-muon events during the LHC beam OFF period. In addition, the CMS collaboration \cite{CMS:2010yju} measured the charge ratio of cosmic-ray muons in the momentum range from $5$ GeV to $1$ TeV, while ATLAS \cite{ATLAS:2010apf} studied the performance and calibration of its detectors using the cosmic-ray data.

\begin{figure}[htb!]
\centering
\includegraphics[scale=0.4]{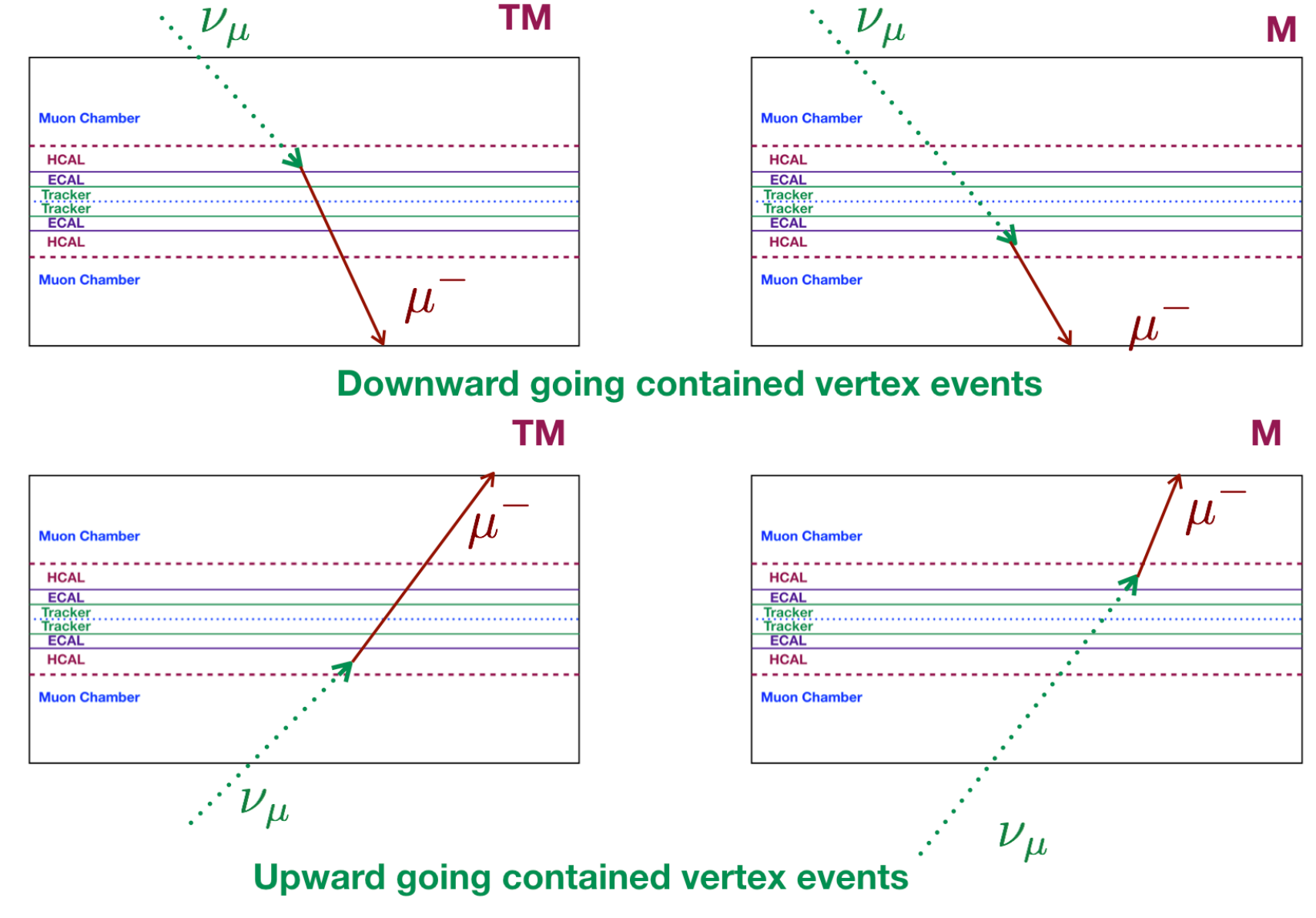}
\caption{{\em Illustration of four distinct categories of downward and upward going contained vertex charged current events induced by atmospheric neutrinos at the ATLAS detector. The {\tt TM} category events will have a charged muon giving hits both in the tracker and muon chamber, while in the {\tt M} category events the muons will only give hits in the muon chamber.}}
\label{fig:ATLAS_signal}
\end{figure}

In the subsequent analyses, we shall look into the details of the characteristic features of the neutrino induced charged current events, and impose event selection conditions to eliminate cosmic-ray muon events at ATLAS in Sec.~\ref{sec:categories}. Although the ATLAS detector sits only at a depth of about 100 m below the ground, in which a large flux of cosmic ray muons penetrate, it is possible to efficiently eliminate the cosmic muon backgrounds by using certain distinct topological features of the events as well as the timing information of the muon chambers. Subsequently, we shall compute the negative and positive muon charge ratio induced by neutrinos and anti-neutrinos as a function of the muon energy at ATLAS, for both contained vertex and external upward going rock muon events in Secs.~\ref{sec:contained} and \ref{sec:upward}, respectively. {\em Although we have focussed on the ATLAS detector throughout this study as a benchmark for large collider detectors, the CMS detector at the LHC can also be used with comparable reach for studying the external upward-going rock muon events.} CMS has a comparable effective area, and an excellent tracker and muon chamber system, which are the key requirements for studying the neutrino induced rock-muons. Due a lower fiducial mass of its core hadron calorimeter, obtaining sufficient yield for the contained-vertex events at CMS will be difficult given the expected exposure time for neutrino physics studies. We summarize our results in Sec.~\ref{sec:summary}, and provide additional details in two Appendices.

\section{Event Categories}
\label{sec:categories}
In Fig~\ref{fig:ATLAS_signal}, we illustrate the four distinct categories of downward and upward going contained vertex charged current events induced by atmospheric neutrinos at the ATLAS detector. The {\tt TM} category events will have a charged muon giving hits both in the tracker and muon chamber, while in the {\tt M} category events the muons will only give hits in the muon chamber. For low-energy muons, we require both tracker and muon chamber hits. As they may not give hits in several layers of the muon chamber, the tracker hits are necessary to successfully reconstruct its trajectory, using the bending of which we can determine the muon's momentum and electric charge. This requirement reduces the low-energy contained events, which is the dominant fraction of such events, by $\sim 50\%$. This stems from the fact that approximately $50\%$ of the events will correspond to {\tt TM} topologies, out of the four topologies shown in Fig~\ref{fig:ATLAS_signal}. This is based on the assumption that both upward and downward going neutrino fluxes are approximately the same. We further assume that the probability of an upward or downward going neutrino interacting in the lower hemisphere is the same, and correspondingly for the interaction probability in the upper hemisphere. The  minimum muon energy considered in our study is 3 GeV, which is necessary to obtain a signature in the muon chamber, after the muon successfully crosses the HCAL without losing all of its energy~\cite{ATLAS:2020ofx}. For higher energy muons (taken to be of energy greater than 10 GeV in our analysis), hits only in the muon chamber are sufficient, as they will give hits on several layers in that part of the detector, and at the same time will have a sufficiently large lever arm for an accurate momentum and charge measurement (recall that the width of the muon chamber on both sides is 7 m each). We further require the muons to be in the central region $|\eta| \leq  2.5 $, but no conditions have been imposed on the azimuthal angles. 

\begin{figure}[htb!]
\centering
\includegraphics[scale=0.4]{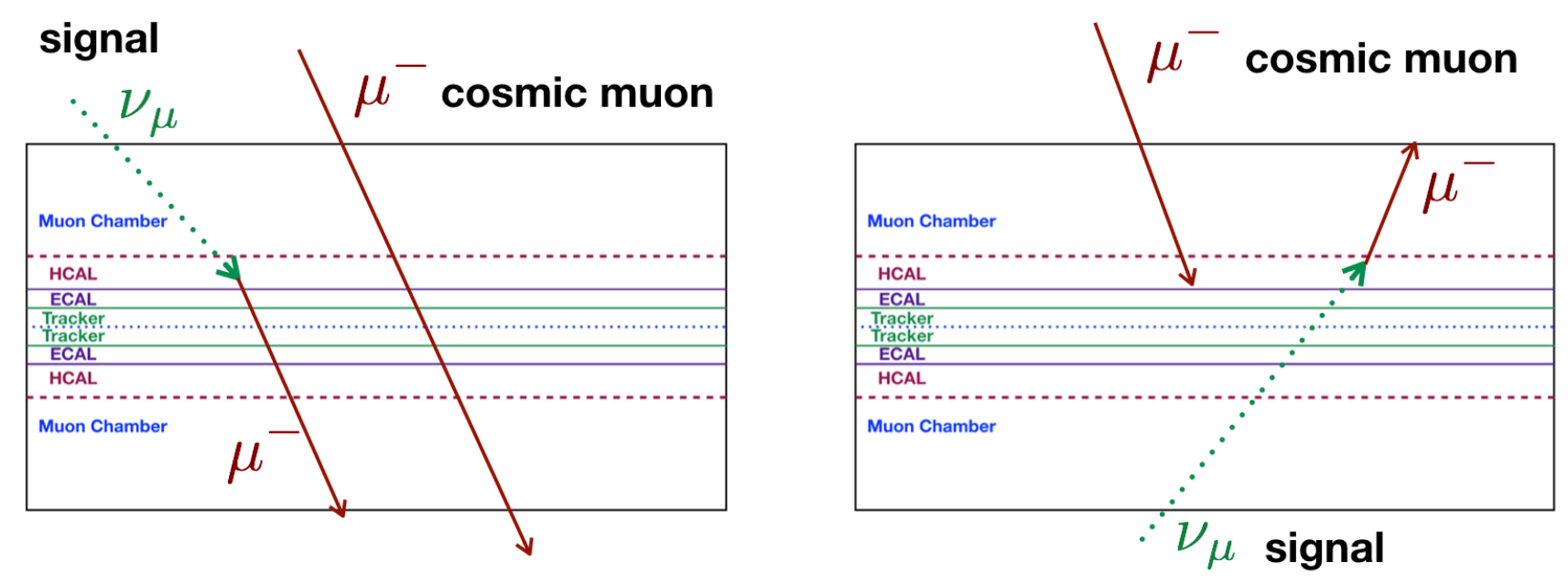}
\includegraphics[scale=0.4]{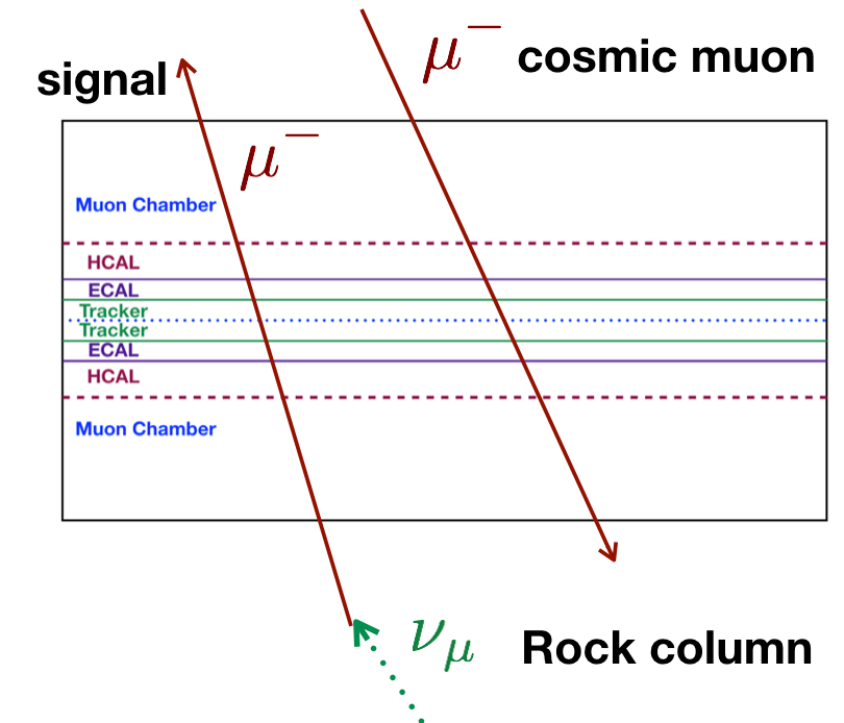}
\caption{{\em Illustration of downward-going cosmic-ray muon events that might mimic corresponding contained vertex charged muon events (Top Panel), and external upward-going charged muon events (Bottom Panel) induced by atmospheric neutrinos at the ATLAS detector, and in the rock column below the detector, respectively. The events that might mimic each other have been shown side-by-side for illustration only.}}
\label{fig:ATLAS_cosmics}
\end{figure}

As illustrated in the top panel of Fig.~\ref{fig:ATLAS_cosmics}, down-going cosmic ray muons cannot mimic down-going contained vertex atmospheric neutrino signal --- cosmic muons first necessarily hit the upper muon chamber, while the signal muon is generated 7 m deeper in the detector at the HCAL. Furthermore, for the upward going contained-vertex signal with no hits at tracker first (i.e., produced at the upper HCAL), timing information is necessary from the resistive plate chamber (RPC) plates of the muon spectrometer. The cosmic muons hit the top-most layer earliest, and conversely for the signal muons. The timing resolution of RPC's is around 1.5 ns \cite{ATLAS:1997ad}, and a muon travels the 7 m muon chamber in about 23 ns, and the full ATLAS width of 22 m in about 73 ns. Therefore, given the timing resolution of the RPCs, we can determine the upward or downward going nature of the muons by checking the time stamps of hits at different layers of the muon chamber. The upward going outside rock muon events can look more similar to cosmic ray muons, as illustrated in Fig.~\ref{fig:ATLAS_cosmics} (Down Panel), and in this case we need to solely rely on the timing information to eliminate the cosmic ray muon backgrounds. Given the sufficient time interval of at least 23 ns or more available to impose the timing requirement, in our study we shall assume a $100\%$ efficiency for this requirement in rejecting cosmic ray muons and in retaining the signal events. A detailed GEANT based analysis can pin down the exact efficiency factor, which is beyond the scope of our study. We further note that although we have assumed that the RPCs cover the full fiducial area of the muon detectors, for ATLAS they span a pseudo-rapidity range of $-1.05 < \eta < 1.3$~\cite{RPC_Table}. For the remaining complementary region, Thin Gap Chambers (TGC) can be utilized, which cover a region $1.05 < |\eta| < 2.7$~\cite{RPC_Table}, and have a similar timing resolution.

Let us now recall the relevant neutrino-nucleon scattering rates, focussing in particular on the difference between neutrino and anti-neutrino cross-sections. There are three distinct types of neutrino-nucleon scattering processes that can take place~\cite{Thomson:2013zua, Formaggio:2012cpf}. At low momentum transfers, there is a quasi-elastic process in which the nucleon changes type but does not break up, with $\nu_\mu + n \rightarrow \mu^- + p$. At slightly higher neutrino energies of the order a few GeV, resonant inelastic processes such as the following have a significant rate: $\nu_\mu + n \rightarrow \mu^- + \Delta^+ \rightarrow \mu^- + p + \pi^0$. Finally, at still higher energies, with order $\sim 4$ GeV for neutrinos and $\sim 8$ GeV for anti-neutrinos, the neutrino interactions begin to be dominated by the neutrino deep inelastic scattering (DIS) processes: ${\nu}_{{\mu}} + N (n,p) \rightarrow \mu^- + X$ and $\overline{\nu}_{{\mu}} + N (n,p) \rightarrow \mu^+ + X$~\cite{Formaggio:2012cpf}. In all these processes, the neutrino cross-sections are larger than anti-neutrino cross-sections. 

The charged current neutrino-nucleon deep inelastic scattering rates, which are the most relevant processes for the neutrino energies of our interest in the ATLAS analysis, can be approximately given for neutrinos by:
\begin{align}
\frac{d\sigma^i}{dE_\mu} \simeq \frac{2 m_p G^2_F}{\pi}\left(a^i+b^i \frac{E^2_\mu}{E^2_\nu}\right)
\label{eq:cccross}
\end{align}      
with $i=(p,n)$, and $a^{p,n}=(0.15,0.25)$, $b^{p,n} =(0.04,0.06)$ and $a^{n,p}_{\bar{\nu}}=b^{p,n}_\nu$, $b^{n,p}_{\bar{\nu}}=a^{p,n}_\nu$, where $p$ and $n$ refer to protons and neutrons, respectively.  Here, $m_p$ is the mass of proton and $G_F$ is the Fermi constant. If we average over the proton and neutron scattering rates, for an isoscalar target, we can write the following combined relations~\cite{Thomson:2013zua}:
\begin{equation}
\frac{d\sigma_{\nu N \rightarrow \mu^-X}}{dE_\mu} \simeq \frac{G_F^2 m_N}{\pi} \left[0.4 + 0.1 \frac{E_\mu^2}{E_\nu^2}\right],
\end{equation}
and for anti-neutrinos by
\begin{equation}
\frac{d\sigma_{\overline{\nu} N \rightarrow \mu^+X}}{dE_\mu} \simeq \frac{G_F^2 m_N}{\pi} \left[0.4 \frac{E_\mu^2}{E_\nu^2} + 0.1 \right].
\end{equation}
As we can see from the above two expressions, the $\frac{d\sigma_{\nu N \rightarrow \mu^-X}}{dE_\mu}$ rate is dominated by the quark contribution of $0.4 \frac{G_F^2 m_N}{\pi}$, while the anti-quark contribution is helicity suppressed due to the left-handed chirality of neutrinos participating in weak interactions. Conversely, for $\frac{d\sigma_{\overline{\nu} N \rightarrow \mu^+X}}{dE_\mu}$, the quark contribution is helicity suppressed, leading to a smaller rate. The helicity suppression is lifted as the charged muon energy approaches the incident neutrino energy, in which limit both the cross-sections become the same. 

\begin{figure}[htb!]
\centering
\includegraphics[scale=0.45]{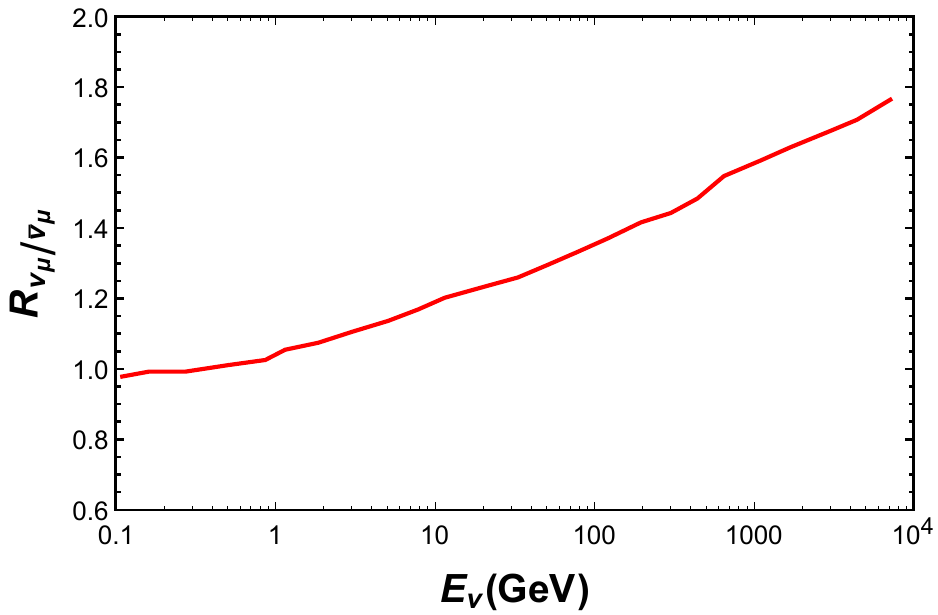}
\hspace{0.3cm}
\includegraphics[scale=0.5625]{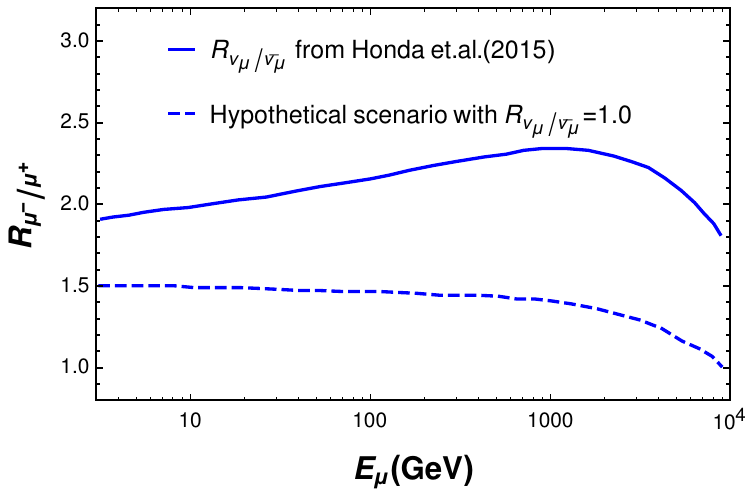}
\caption{{\em (Left)The flux ratio $R_{\nu_\mu/ \overline{\nu}_\mu}$ for atmospheric muon neutrinos as a function of $E_\nu$, with the zenith angle averaged fluxes taken from Honda et al.(2015)~\cite{Honda:2015fha}, for the Kamioka site. (Right) The expected negative and positive charge muon ratio $R_{\mu^-/ \mu^+}$ for contained vertex events at the ATLAS detector as a function of the muon energy $E_\mu$, assuming $100\%$ efficiency, with the actual flux ratio $R_{\nu_\mu/ \overline{\nu}_\mu}$ (solid line), and for a hypothetical scenario in which $R_{\nu_\mu/ \overline{\nu}_\mu}=1$ (dashed line) for all energies. }}
\label{Fig:expected}
\end{figure}

In Fig.~\ref{Fig:expected} (left column), we show the flux ratio $R_{\nu_\mu/ \overline{\nu}_\mu}$ for atmospheric muon neutrinos as a function of $E_\nu$, with the zenith angle averaged fluxes taken from Honda et al.(2015)~\cite{Honda:2015fha}, for the Kamioka site~\footnote{Since there is no atmospheric neutrino flux computation available for the CERN site, we have used the flux predictions for Kamioka, which is at a similar latitude as the CERN Geneva site~\cite{Latitude}.}. In the right column of the same figure, we show the expected negative and positive charge muon ratio $R_{\mu^-/ \mu^+}$ for contained vertex events at the ATLAS detector as a function of the muon energy $E_\mu$, assuming $100\%$ efficiency, with the actual flux ratio $R_{\nu_\mu/ \overline{\nu}_\mu}$ (solid line), and for a hypothetical scenario in which $R_{\nu_\mu/ \overline{\nu}_\mu}=1$ (dashed line) for all energies. The $R_{\nu_\mu/ \overline{\nu}_\mu}=1$ line shows the impact of the difference between neutrino and anti-neutrino nucleon cross-sections that we discussed above. The solid line is a convolution of the actual flux ratio $R_{\nu_\mu/ \overline{\nu}_\mu}$ and the DIS cross-section differences, and the turnover at energies above a few TeV is expected since the DIS events in this region of muon energy are mostly coming from neutrinos of similar energy, where the DIS cross-sections for neutrinos and anti-neutrinos become similar. 

\section{Contained vertex Events}
\label{sec:contained}
We now discuss the event rates expected at the ATLAS detector --- for the contained vertex type events, as well as for the upward-going outside muon events. For a given flux of muon (anti-)neutrinos with energy $E_\nu$, producing (anti-)muons with energy $E_\mu$ at the detector, the contained vertex muon flux is computed as:   
\begin{align}
\frac{dN_\mu}{dE_\mu dV dt} = \int d\Omega \int^\infty_{E_\mu} dE_\nu \frac{d\Phi_\nu}{dE_\nu}\left[\frac{d \sigma^p}{dE_\mu} n_p + (p\rightarrow n)\right]
\label{eq:conmuon}
\end{align}  
where, $n_i$ ($i=p,n$) are the number densities of nucleons present in the fiducial detector volume. $d\Omega$ is the differential solid angle subtended by a patch of the sky w.r.t the axis, perpendicular to the plane of ATLAS detector. Since the analysis is performed with a neutrino flux averaged over the zenith angles, the angular acceptance of the detector affecting the event rates, can only be addressed by a detailed detector simulation based on GEANT, which is beyond the scope of this work. We have integrated over the $4\pi$ solid angle for computing the total number of expected events. The impact parameter of the incident neutrinos in the HCAL has not been considered -- we have assumed that the fiducial mass of the HCAL is uniformly distributed over its fiducial volume.

In Table~\ref{tab:contained}, we show the contained vertex events for $\mu^-$ and $\mu^+$, along with their ratios at the ATLAS detector in the central region $|\eta| \leq  2.5 $ for different energy ranges, with ($4$ kiloton $\times 1000$ days) $\sim 11$ kiloton-year exposure, satisfying the event selection criteria mentioned in the category column for different energy ranges. We have used $95\%$ efficiency for muon tagging and charge-identification of a muon for the entire energy range. This choice stems from the medium muon selection criteria, as shown in Refs.\cite{ATLAS:2016lqx,ATLAS:2020auj}, however, a loose selection criteria may enhance the efficiency further. We have conservatively required a minimum muon energy of 10 GeV , which is adequate to ensure sufficient number of hits in the muon chamber to allow reconstruction.  As we see from this table, first of all, the total number of events after all selection cuts is significant enough to carry out a contained-vertex event search with $1000$-live days of neutrino study. We obtain $60 ~\mu^-$ events, and $30 ~\mu^+$ events with $ E_\mu\geq 3$ GeV. As we see from this table, there is an energy dependence in $R_{\mu^-/ \mu^+}$, with a higher ratio at higher energies. The value of $R_{\mu^-/ \mu^+}$ averaged over all energies is obtained to be $R_{\mu^-/ \mu^+}=2.00^{+0.53}_{-0.39}$, where we have indicated the expected $68\%$ C.L. statistical error computed by using the ratio distribution of two random numbers following the Poisson distribution \cite{James:1980my, Helene:1983ph}. To note, the ratio distribution is found to be asymmetric around the most probable value, which results into an asymmetric error bars for $R_{\mu^-/\mu+}$. For details please see Appendix.\ref{app:A}. 
\begin{table}[htb!]
\centering
\hspace{-1.5cm}
\begin{tabular}{|l|l|l|l|l|}
\hline
\begin{tabular}[c]{@{}l@{}} Energy  \end{tabular} & \begin{tabular}[c]{@{}l@{}}$N_{\mu^-}$  \end{tabular} & \begin{tabular}[c]{@{}l@{}}$N_{\mu^+}$ \\ \end{tabular} & \begin{tabular}[c]{@{}l@{}}$N_{\mu^-}/N_{\mu^+}$\end{tabular} & \begin{tabular}[c]{@{}l@{}} Category \\\end{tabular} \\ \hline
$3\leq E_\mu \leq 10$ GeV  & $31$  & $16$ & $1.82$ & Only TM  \\ \hline
$5\leq E_\mu \leq 10$ GeV & $13$ & $ 7$ & $1.86$ & Only TM \\ \hline
$E_\mu>10$ GeV & $29$ & $14$ & $2.07$ & TM $\&$ M  \\ \hline
$E_\mu>20$ GeV & $14$ & $7$ & $2.00$ & TM $\&$ M \\ \hline
Total: $ E_\mu\geq 3$ GeV & $60$ & $30$ & $2.00$ & \\\hline
\end{tabular}
\caption{{\em Contained vertex events for $\mu^-$ and $\mu^+$, along with their ratios at the ATLAS detector in the central region $|\eta| \leq 2.5 $ for different energy ranges, with $11$ kiloton-year exposure, satisfying the event selection criteria mentioned in the category column for different energy ranges.}}
\label{tab:contained}
\end{table}

\begin{figure}[htb!]
\centering
\includegraphics[scale=0.38]{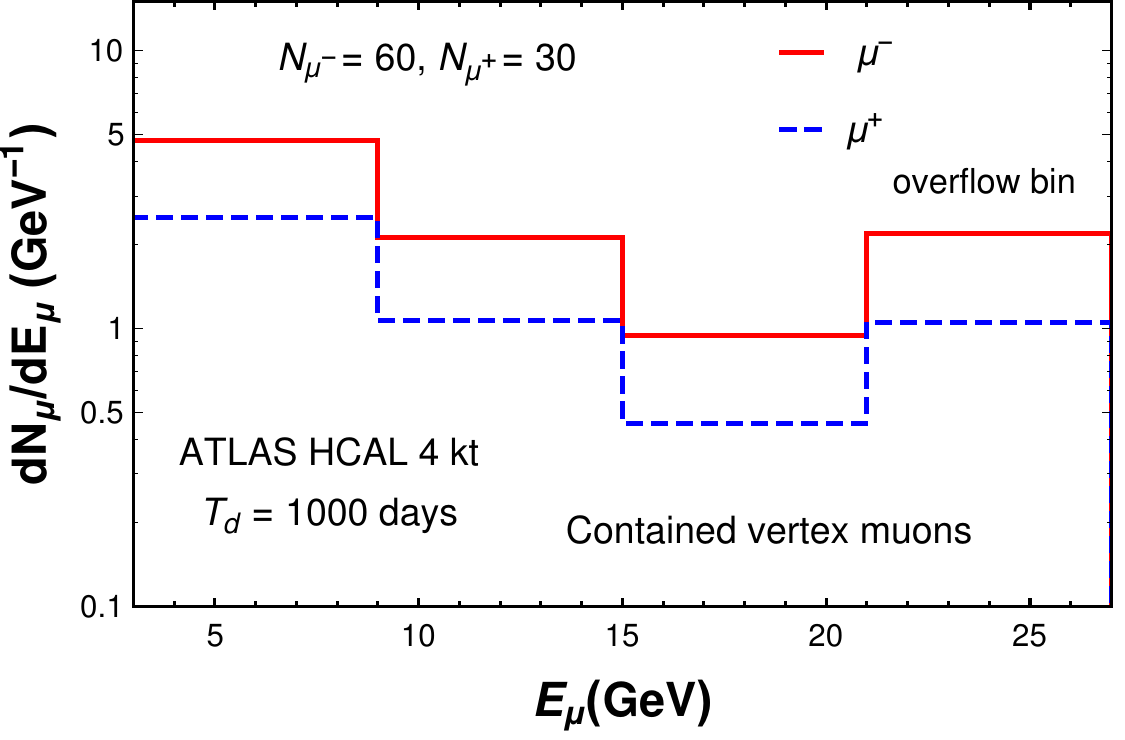}
\includegraphics[scale=0.37]{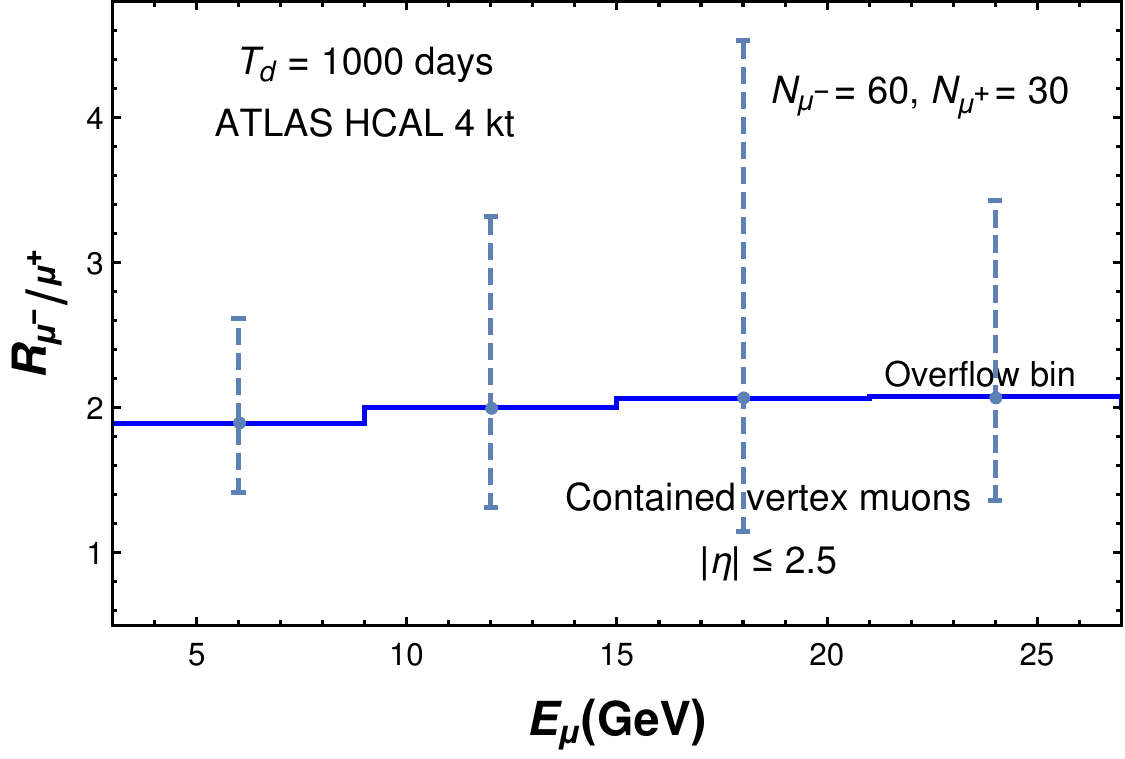}
\caption{{\em (Left) Differential energy distribution of contained vertex events for $\mu^-$ (solid red line) and $\mu^+$ (dashed blue line), along with (Right) their ratio $R_{\mu^-/ \mu^+}$ at the ATLAS detector in the central region $|\eta| \leq 2.5 $ as a function of the muon energy $E_\mu$, with $11$ kiloton-year exposure, satisfying the event selection criteria discussed in the text.}}
\label{Fig:diff_contained}
\end{figure}

In Fig.~\ref{Fig:diff_contained}, we show (left panel) the differential energy distribution of contained vertex events for $\mu^-$ (solid red line) and $\mu^+$ (dashed blue line), along with (right panel) their ratio $R_{\mu^-/ \mu^+}$ at the ATLAS detector in the central region $|\eta| \leq  2.5 $ as a function of the muon energy $E_\mu$, with $11$ kiloton-year exposure, satisfying the event selection criteria described earlier. $14$ $\mu^-$ and $7$ $\mu^+$ events with $E_\mu>20$ GeV are shown in the \textit{overflow} bin of these figures at the right edge of the plot. The energy dependence of $R_{\mu^-/ \mu^+}$ has been shown in the right panel of Fig.\ref{Fig:diff_contained}. While we have given the error estimate of $R_{\mu^-/ \mu^+}$ above considering all events with $E_\mu>3$ GeV, we have also shown the error bars in the individual energy bins, which become larger for the higher energy bins due to smaller number of contained-vertex events -- this aspect will be improved upon below for the external rock muon events, with a much larger event count at ATLAS. However, in order to increase the contained event count, the LHC collaboration may look into atmospheric neutrino events with a primary vertex
veto (no collision), during the beam ramp up and ramp down phases, when the pile-up may be low \cite{Steerenberg}, provided the detectors can be kept operational during these phases.

\section{External upward-going muon events}
\label{sec:upward}
We next discuss the flux of external upward-going muons coming from the charged current muon neutrino interactions in the rock surrounding the detector material. In this case, subsequent to their production, the muons lose energy on the way to the detector, as parametrized by the stopping potential of muons on rock~\cite{Lohmann:1985qg, Groom:2001kq}. For a neutrino with energy $E_\nu$,  producing a muon with initial energy $E^0_\mu$, and final energy $E_\mu$ as it reaches the detector, the flux can be expressed as~\cite{Gaisser_Book, Erkoca:2009by, Covi:2009xn}:
\begin{align}
\frac{dN}{dE_\mu dA dt}=\int d\Omega\int^{\infty}_{E_\mu} dE_\nu \int^{R_\mu(E_\mu,E_\nu)}_0 dr~ e^{\beta\rho r} P_{\rm sur} (E_\mu, r) \frac{d\Phi_\nu}{dE_\nu}\left[ \frac{d\sigma^p}{dE^0_\mu} n_p ~ +(p\rightarrow n)\right].
\label{eq:thmuon}
\end{align}
The initial energy $E^0_\mu$ of the muons is related to the energy $E_\mu$ as it reaches the detector by the following:
\begin{align}
E^0_\mu = E_\mu e^{\beta \rho r}+ \frac{\alpha}{\beta}\left(e^{\beta \rho r}-1\right),
\label{eq:muenergy}
\end{align}   
where, we have used the approximate energy independent coefficients for muon energy loss in matter: $\alpha =2.3 \times 10^{-3} ~\rm GeV cm^2~ gm^{-1}$ and $\beta = 4.4\times 10^{-6}~ \rm cm^2 ~gm^{-1}$~\cite{Erkoca:2009by, Covi:2009xn, Lohmann:1985qg, Groom:2001kq}. Here, $\rho$ is the density of matter surrounding the detector, and assuming it to be rock, we take $\rho =2.65 ~\rm gm ~cm^{-3}$~\cite{Lohmann:1985qg, Groom:2001kq}. The effective range of muons, $r$, is defined as the distance from the 
detector volume to the primary interaction vertex of the neutrinos outside the detector. The maximum possible value of the effective range, $R_\mu$, of a muon can be written as a function of $E_\mu$ measured at the detector and the original neutrino energy as follows~\cite{Erkoca:2009by}:
\begin{align}
R_\mu(E_\mu,E_\nu)= \frac{1}{\rho \beta}
  \log\left[\frac{\alpha + \beta E_\nu}{\alpha + \beta E_\mu}\right] .
\end{align}
In addition to the energy loss in rock, muons may also decay while going through the medium before reaching the detector. This effect is taken into account via the survival probability given by~\cite{Erkoca:2009by}:
\begin{align}
P_{\rm sur} (E_\mu, r)= \left(\frac{E_\mu}{E^0_\mu}~\frac{\alpha+\beta E^0_\mu}{\alpha+\beta E_\mu}\right)^{\Gamma}
\end{align}
where, $\Gamma = m_\mu/(\alpha L \rho)$, $L (=c \tau = 0.65 ~\rm km)$ being the decay length of muon at rest and $m_\mu$ is the muon mass. There can be neutrino oscillations while propagating through the matter of the earth, which can reduce the flux of muon neutrinos due to its conversion to electron and tau neutrinos. However, the survival probability of muon neutrino for a incident neutrino energy of around 10 GeV is found to be $0.98$ and for $100$ GeV is $0.92$, which implies negligible change in subsequent muon flux. Therefore, we have ignored the effect of neutrino oscillation in matter.\footnote{We have used the analytical expression given in Eq.14 of Ref.\cite{Gandhi:2004bj} to calculate the survival probability of muon neutrino in matter, while taking relevant parameters as \cite{Esteban:2024eli}, $\sin^2 \theta_{13}=0.02$, $\sin^2 \theta_{23}=0.454$, $\Delta m^2_{31}=0.0025 ~{\rm eV^2}$ and $L=9700$ km. }

\begin{table}[htb!]
\centering
\begin{tabular}{|l|l|l|l|}
\hline
\begin{tabular}[c]{@{}l@{}} Energy  \end{tabular} & \begin{tabular}[c]{@{}l@{}}$N_{\mu^-}$  \end{tabular} & \begin{tabular}[c]{@{}l@{}}$N_{\mu^+}$ \\ \end{tabular} & \begin{tabular}[c]{@{}l@{}}$N_{\mu^-}/N_{\mu^+}$\end{tabular}  \\ \hline
$3\leq E_\mu \leq 10$ GeV  & $157$  & $83$ & $1.89$  \\ \hline
$5\leq E_\mu \leq 10$ GeV & $91$ & $48$ & $1.90$   \\ \hline
$E_\mu>10$ GeV & $442$ & $209$ & $2.11$    \\ \hline
$E_\mu>20$ GeV & $350$ & $162$ & $2.16$   \\ \hline
Total: $ E_\mu\geq 3$ GeV & $599$ & $ 292$ & 2.05  \\\hline
\end{tabular}
\caption{{\em External upward-going charged muon events at the ATLAS detector in the central region $|\eta| \leq 2.5 $ with $1000$-live days of neutrino study, for different energy ranges, generated at the rock column below the detector from upward going neutrinos.}}
\label{tab:external}
\end{table}

In Table~\ref{tab:external}, we show the external upward-going charged muon events at the ATLAS detector in the central region $|\eta| \leq 2.5 $ with $1000$-live days of neutrino study, for different energy ranges, generated at the rock column below the detector from upward going neutrinos. As we see from this table, the total number of events after all selection cuts is much larger than the contained vertex events --- this is because of the larger fiducial mass available in the rock column for neutrino nucleon DIS scattering. Since the muon effective range increases in rock with increasing energy, higher energy muons can come from a larger depth. Therefore, correspondingly, higher energy neutrinos have a larger effective rock volume available for interactions, in addition to having a larger DIS cross-section. Because of this fact, even though the flux of atmospheric neutrinos fall with energy as approximately $E^{-2.7}$, the number of produced muons falls with energy much slower. Thus, as far as the total event count is concerned, external upward-going charged muon events are more promising. We obtain $599 ~\mu^-$ events, and $292 ~\mu^+$ events with $ E_\mu\geq 3$ GeV. The energy dependence in $R_{\mu^-/ \mu^+}$ is now milder, with a higher ratio at higher muon energies. This is because neutrinos of a range of energies contribute to this class of events with the same muon energy. The value of $R_{\mu^-/ \mu^+}$ averaged over all energies is obtained to be $R_{\mu^-/ \mu^+}=2.05^{+0.15}_{-0.14}$, where we have indicated the expected $68\%$ C.L. statistical error~\cite{James:1980my, Helene:1983ph}, which is  smaller than in the contained vertex events scenario. If instead of the Honda flux, one adopts the Bartol flux~\cite{Barr:2004br}, we obtain the expected ratio to be $R_{\mu^-/\mu^+}=1.95^{+0.17}_{-0.15}$ at $68\%$ C.L.

\begin{figure}[htb!]
\centering
\includegraphics[scale=0.45]{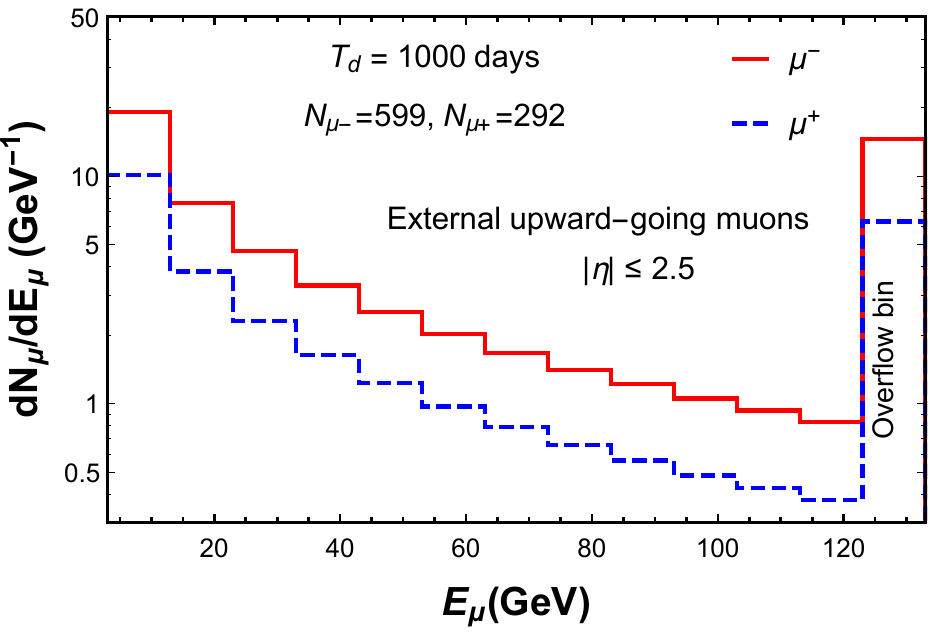}  
\includegraphics[scale=0.45]{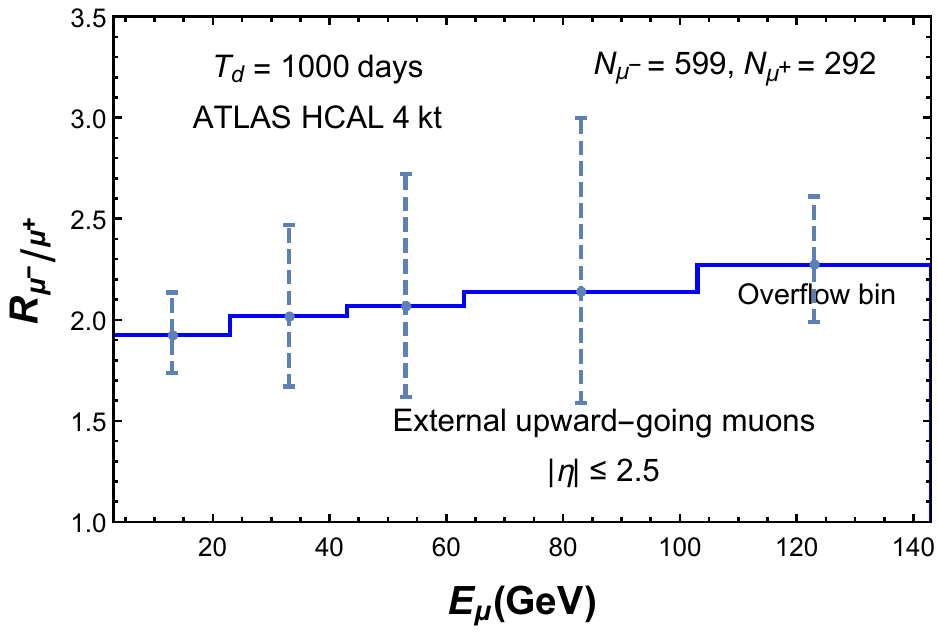}
\caption{{\em (Left) Differential energy distribution of external upward-going events generated at the rock column below the detector from upward going neutrinos for $\mu^-$ (solid red line) and $\mu^+$ (dashed blue line), along with (Right) their ratio $R_{\mu^-/ \mu^+}$ at the ATLAS detector in the central region $|\eta| \leq 2.5 $ as a function of the muon energy $E_\mu$, with $1000$-live days of neutrino study, satisfying the event selection criteria. The error bars indicated for different $R_{\mu^-/ \mu^+}$ bins correspond to the expected $68\%$ C.L. statistical error.}}
\label{Fig:diff_external}
\end{figure}

In Fig.~\ref{Fig:diff_external}, we show (left panel) the differential energy distribution of external upward-going events generated at the rock column below the detector from upward going neutrinos for $\mu^-$ (solid red line) and $\mu^+$ (dashed blue line), along with (right panel) their ratio $R_{\mu^-/ \mu^+}$ at the ATLAS detector in the central region $|\eta| \leq 2.5 $ as a function of the muon energy $E_\mu$, with $1000$-live days of neutrino study, satisfying the event selection criteria. As discussed in detail in the context of Table~\ref{tab:external} above, the  number of events is here is much larger, and we can study $R_{\mu^-/ \mu^+}$ upto an energy of around 100 GeV, within reasonable errors, as indicated by the $68\%$ C.L. error bars, computed following Refs.~\cite{James:1980my, Helene:1983ph}. With these events, a $R_{\mu^-/ \mu^+}$ ratio of around 1.5, which follows from a $R_{\nu_\mu/ \overline{\nu}_\mu}$ ratio of 1 (see Fig.~\ref{Fig:expected}), may also be excluded at $68\%$ C.L. upto a muon energy of around 60 GeV, thus probing the asymmetry in the atmospheric neutrino and anti-neutrino flux with the ATLAS data. We note that the ratio of contained vertex events and the external upward-going events is different in our analysis, as compared to the one reported in the atmospheric neutrino literature, for example by the Super-Kamiokande collaboration. This difference is explained in Appendix \ref{app:B}. 

Before concluding, we present a numerical estimate of the upward-going cosmic muon background to the upward-going neutrino events. As we discussed above, muons lose a considerable energy while travelling through the rock inside the earth. For example, since we need at least a 3 GeV muon in the ATLAS detector, the incident energy of the muon on earth should then be at least 1.1 TeV, if it travels 1 km in the rock before reaching the detector. Therefore, it is clear that ``upward-going" muons coming only from zenith angles close to $90^\circ$ will have some chance to reach the detector before losing most of its energy. Since the ATLAS detector sits at a depth of $100$ meters below the surface of the earth, if we take the zenith angle of the incident muon to be $89^\circ$, it needs to travel a minimum length of around $5.72$ km through earth. Therefore, to be registered in the muon chamber after travelling through the detector, it needs to have an incident energy greater than or equal to $4.13\times10^5$ GeV. We have estimated that the number of cosmic muons (using the flux given in Ref.\cite{Gaisser_Book}) with such energy that will be incident on the ATLAS detector with a 1000 live-days of running is $7.54 \times 10^{-4}$, which is negligible.

\section{Summary}
\label{sec:summary}
To summarize our study, as is well-known, the ratio of atmospheric muon neutrinos and anti-neutrinos is an important quantity, for which there is still a large uncertainty in the prediction of the different flux models. Therefore, it is important to be able to directly measure this quantity in neutrino physics experiments, as a function of neutrino energy. However, most neutrino detectors do not distinguish between muon neutrinos and anti-neutrinos. Magnetized detectors can discriminate on an event-by-event basis between (anti-)neutrino induced events by measuring the electric charge of the (anti-)muon: MINOS experiment had such a detector, which reported a measurement of this ratio combining all events, but did not report the energy dependence of it. The large collider detector ATLAS at CERN LHC can be used for this purpose during the periods when the LHC beams are not in circulation -- it is sufficiently heavy for neutrino physics (the hadron calorimeter weighs 4 kilotons), and finely instrumented to reject cosmic ray muon backgrounds. 

We have discussed in detail the distinction between the cosmic ray muons and the signal muons for different categories of events, suggesting suitable event selection criteria to eliminate the cosmic ray muon backgrounds. While contained-vertex events are the most striking, upward-going outside events are larger in number at ATLAS for muon energies greater than 3 GeV, due to the larger fiducial mass available from the rock-column in the earth below the detector. For both these categories, we estimated the event rates expected after employing suitable event selection criteria necessary to eliminate the cosmic ray muon background. We find that ATLAS, with a 1000-live days of neutrino physics, spanned over a period of 10--15 years especially during the winter months, can accumulate sufficient events to study the energy dependence of the charged muon ratio induced by atmospheric neutrinos and anti-neutrinos. With this exposure, we obtain $60~\mu^-$ and $30~\mu^+$ contained vertex events, and $599~\mu^-$ and $292~\mu^+$ external upward-going events, after imposing the necessary selection criteria. 

The CMS detector at the LHC can also be used with comparable reach for studying the external upward-going rock muon events. {\em Therefore, given its possible impact in improving atmospheric neutrino flux models, and consequently in neutrino physics, the ATLAS and CMS detectors should be utilized for studying atmospheric muon neutrinos and anti-neutrinos, and we hope that further detailed detector-level analyses will be carried out by the ATLAS and CMS collaborations in this regard.} In addition to such atmospheric neutrino studies, if an excess of neutrino signal over the expected atmospheric and astrophysical backgrounds is observed in any of the large dedicated neutrino telescopes, {\em the ratio of neutrinos and anti-neutrinos in the excess may be determined using ATLAS or CMS in subsequent analyses.  This can provide a vital clue in determining the origin of the excess}, possibly in physics beyond the standard model scenarios, such as leptogenesis or asymmetric dark matter, in which the lepton-antilepton symmetry breaking can manifest in a neutrino-antineutrino signal asymmetry~\cite{Feldstein:2010xe, Fukuda:2014xqa}.

\section*{Acknowledgements}
We are grateful to Raj Gandhi for several valuable inputs on different neutrino experiments. We are very much thankful to Swagata Mukherjee and Sunanda Banerjee for discussions on trigger, event selection and muon identification at ATLAS/CMS, to Atri Bhattacharya for detailed discussions on the muon (anti-)neutrino fluxes in atmospheric neutrinos, and to Mehedi Masud for help with neutrino oscillations in matter. We also thank Suchandra Dutta, Subir Sarkar, Biplob Bhattacherjee, Rohan Pramanick, Shankha Banerjee, and D.~Indumathi for many helpful discussions. We thank Satyaki Bhattacharya for useful inputs regarding the statistical analysis. 

\appendix
\section{Statistical error analysis of the muon-antimuon ratio}
\label{app:A}
The number of contained-vertex events per bin is rather small so that the Gaussian approximation fails although we can use the same for the total number of events. For the bin-wise error bar on $R_{\mu^-/\mu^+}$, we use the following probability distribution function as reported in Ref.\cite{Helene:1983ph}.

\begin{align}
f(R_{\mu^-/\mu^+})= \frac{(N_{\mu^-}+N_{\mu^+}+1)!}{N_{\mu^-}!~ N_{\mu^+}!} \frac{R_{\mu^-/\mu^+}^{N_{\mu^-}}}{(1+R_{\mu^-/\mu^+})^{N_{\mu^-}+N_{\mu^+}+2}}
\label{eq:dist}
\end{align}
To check the viability of the above distribution function, we have numerically determined the probability distribution function of the ratio of two random variables taken from two independent Poisson distribution. Our numerically generated distribution is well approximated by Eq.\ref{eq:dist}. As the distribution is asymmetric, the confidence interval around the central value is not unique. However, we have used an ``one sigma equivalent" around the central value of $R_{\mu^-/\mu^+}$ to show the statistical error at $68\%$ C.L. In particular, we have found asymmetric error bars ($\Delta R_+$, $\Delta R_-$) demanding the probability to be 0.34, at each side of the central value ($R_0$) separately, i.e.
\begin{align}
 \int^{R_0+\Delta R_+}_{R_0} f(R_{\mu ^-/\mu^+}) dR_{\mu ^-/\mu^+} =0.34,~~ \int^{R_0}_{R_0-\Delta R_-} f(R_{\mu ^-/\mu^+}) dR_{\mu ^-/\mu^+} =0.34
\label{eq:stat}
\end{align}
With these definitions we have shown the bin-wise statistical error in the muon-antimuon ratio for the contained (external upward-going) muon events in Fig.\ref{eq:conmuon} (Fig.\ref{Fig:diff_external}).

\section{Comparison of event rates at the ATLAS and Super-Kamiokande detectors}
\label{app:B}

\begin{figure}[htb!]
\centering
\includegraphics[scale=0.6]{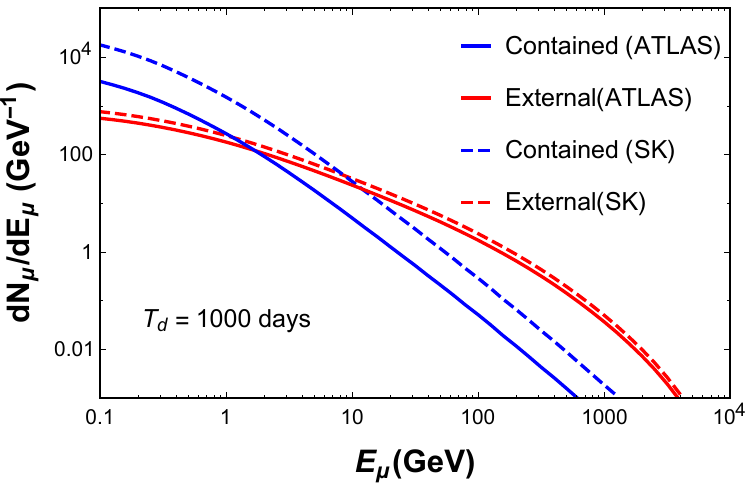}  
\caption{{\em Comparison of the charged current $(\nu_\mu+\overline{\nu}_\mu)$ event rates at the Super-Kamiokande (SK) and ATLAS detectors with a $1000$-live days of exposure, for both contained-vertex type and external upward-going events, without any selection cuts, in the muon energy range of 0.1 GeV to 10 TeV.}}
\label{Fig:ATLAS_vs_SK}
\end{figure}

As mentioned in the main text, the ratio of contained vertex events and the external upward-going events is different in our analysis, as compared to the one reported in the atmospheric neutrino literature, for example by the Super-Kamiokande collaboration. The primary reason for this is twofold. The number of contained-vertex events is proportional to the detector fiducial mass, which for SK is $\sim 5$ times larger than the ATLAS HCAL. On the other hand, the number of external upward-going events is proportional to the effective area of the detector exposed to the rock muons, which is similar for the two experiments, with SK having around $1200 ~m^2$ and ATLAS around $880 ~m^2$, making SK have an effective area bigger by a factor of 1.36. Thus we expect the number of external rock-muon events to be similar in both SK and ATLAS in the full energy range, while the number of contained vertex events should be much larger in SK compared to ATLAS. Both these features are seen in Fig.~\ref{Fig:ATLAS_vs_SK}, where we have shown a comparison of the charged current $(\nu_\mu+\overline{\nu}_\mu)$ event rates at the Super-Kamiokande and ATLAS detectors with a $1000$-live days of exposure, for both contained-vertex type and external upward-going events, without any selection cuts, in the muon energy range of 0.1 GeV to 10 TeV. 

The second reason is the different event selection criteria employed in the two experiments. For ATLAS, we have only included events with  muon energy $E_\mu > 3$ GeV. As we can see from Fig.~\ref{Fig:ATLAS_vs_SK}, for $E_\mu>3$ GeV, ATLAS has many more external rock muon events than contained vertex ones, as we also have found in our numerical analysis above. Since SK uses a much lower threshold of $100$ MeV for muon energy, for $E_\mu \lesssim 10$ GeV, they have a much larger number of contained vertex events. These two reasons combined together clearly explain why the ratio of the two types of events at ATLAS found in our analysis is oppositely ordered than in SK analysis~\footnote{We thank D.~Indumathi for asking us about this comparison between SK and ATLAS.}. We have cross-checked our results on both categories of events with the existing literature for detectors such as Super-Kamiokande. Furthermore, the event selection criteria we have used to eliminate cosmic ray muon events leads to a further reduction of around $50\%$ contained vertex events, but no such reduction happens for the typically higher energy external upward-going events.


\begin{thebibliography}{99}
\bibitem{Gaisser_Book}
T.K.~Gaisser,
Cosmic Rays and Particle Physics, Cambridge University Press, 1990.

\bibitem{Volkova:1980sw}
L.~V.~Volkova,
``Energy Spectra and Angular Distributions of Atmospheric Neutrinos,''
Sov. J. Nucl. Phys. \textbf{31}, 784-790 (1980).


\bibitem{Honda:2015fha}
M.~Honda, M.~Sajjad Athar, T.~Kajita, K.~Kasahara and S.~Midorikawa,
``Atmospheric neutrino flux calculation using the NRLMSISE-00 atmospheric model,''
Phys. Rev. D \textbf{92}, no.2, 023004 (2015).

\bibitem{Honda:2011nf}
M.~Honda, T.~Kajita, K.~Kasahara and S.~Midorikawa,
``Improvement of low energy atmospheric neutrino flux calculation using the JAM nuclear interaction model,''
Phys. Rev. D \textbf{83}, 123001 (2011).


\bibitem{SK_Full}
Y.~Ashie\textit{et al.} [Super-Kamiokande],
``Measurement of atmospheric neutrino oscillation parameters by Super-Kamiokande I",
Phys. Rev. D \textbf{71}, 112005 (2005).

\bibitem{Battistoni:2003ju}
G.~Battistoni, A.~Ferrari, T.~Montaruli and P.~R.~Sala,
``High-energy extension of the FLUKA atmospheric neutrino flux,''
[arXiv:hep-ph/0305208 [hep-ph]].

\bibitem{Honda:2004yz}
M.~Honda, T.~Kajita, K.~Kasahara and S.~Midorikawa,
``A New calculation of the atmospheric neutrino flux in a 3-dimensional scheme,''
Phys. Rev. D \textbf{70}, 043008 (2004).

\bibitem{Barr:2004br}
G.~D.~Barr, T.~K.~Gaisser, P.~Lipari, S.~Robbins and T.~Stanev,
``A Three - dimensional calculation of atmospheric neutrinos,''
Phys. Rev. D \textbf{70}, 023006 (2004).


\bibitem{Super-Kamiokande:2024rwz}
H.~Kitagawa \textit{et al.} [Super-Kamiokande],
``Measurements of the charge ratio and polarization of cosmic-ray muons with the Super-Kamiokande detector,''
Phys. Rev. D \textbf{110}, no.8, 082008 (2024)
doi:10.1103/PhysRevD.110.082008
[arXiv:2403.08619 [hep-ex]].

\bibitem{Yanez:2019bnw}
J.~P.~Y\'a\~nez, A.~Fedynitch and T.~Montgomery,
``Calibration of atmospheric neutrino flux calculations using cosmic muon flux and charge ratio measurements,''
PoS \textbf{ICRC2019}, 881 (2020)
doi:10.22323/1.358.0881
[arXiv:1909.08365 [astro-ph.HE]].

\bibitem{Super-Kamiokande:2015qek}
E.~Richard \textit{et al.} [Super-Kamiokande],
``Measurements of the atmospheric neutrino flux by Super-Kamiokande: energy spectra, geomagnetic effects, and solar modulation,''
Phys. Rev. D \textbf{94}, no.5, 052001 (2016)
doi:10.1103/PhysRevD.94.052001
[arXiv:1510.08127 [hep-ex]].

\bibitem{IceCubeCollaboration:2023wtb}
R.~Abbasi \textit{et al.} [IceCube Collaboration],
``Measurement of atmospheric neutrino mixing with improved IceCube DeepCore calibration and data processing,''
Phys. Rev. D \textbf{108}, no.1, 012014 (2023)
doi:10.1103/PhysRevD.108.012014
[arXiv:2304.12236 [hep-ex]].


\bibitem{Beacom:2003nk}
J.~F.~Beacom and M.~R.~Vagins,
``GADZOOKS! Anti-neutrino spectroscopy with large water Cherenkov detectors,''
Phys. Rev. Lett. \textbf{93} (2004), 171101
doi:10.1103/PhysRevLett.93.171101
[arXiv:hep-ph/0309300 [hep-ph]].

\bibitem{Super-Kamiokande:2022hxq}
K.~Abe \textit{et al.} [Super-Kamiokande],
``Neutron tagging following atmospheric neutrino events in a water Cherenkov detector,''
JINST \textbf{17}, no.10, P10029 (2022)
doi:10.1088/1748-0221/17/10/P10029
[arXiv:2209.08609 [hep-ex]].




\bibitem{Super-Kamiokande:2023ahc}
T.~Wester \textit{et al.} [Super-Kamiokande],
``Atmospheric neutrino oscillation analysis with neutron tagging and an expanded fiducial volume in Super-Kamiokande I\textendash{}V,''
Phys. Rev. D \textbf{109}, no.7, 072014 (2024)
doi:10.1103/PhysRevD.109.072014
[arXiv:2311.05105 [hep-ex]].


\bibitem{MINOS:2008hdf}
D.~G.~Michael \textit{et al.} [MINOS],
``The Magnetized steel and scintillator calorimeters of the MINOS experiment,''
Nucl. Instrum. Meth. A \textbf{596}, 190-228 (2008).

\bibitem{Gaisser:2002zv}
T.~K.~Gaisser and T.~Stanev,
``Charge ratio of muons from atmospheric neutrinos,''
Phys. Lett. B \textbf{561}, 125-129 (2003).



\bibitem{MINOS_Atmospheric}
P.~Adamson \textit{et al.} [MINOS],
``Measurements of atmospheric neutrinos and antineutrinos in the MINOS Far Detector,''
Phys. Rev. D \textbf{86}, 052007 (2012)

\bibitem{ATLAS:2008xda}
G.~Aad \textit{et al.} [ATLAS],
``The ATLAS Experiment at the CERN Large Hadron Collider,''
JINST \textbf{3}, S08003 (2008)

\bibitem{Kopp:2007ai}
J.~Kopp and M.~Lindner,
``Detecting atmospheric neutrino oscillations in the ATLAS detector at CERN,''
Phys. Rev. D \textbf{76}, 093003 (2007).

\bibitem{ATLAS:2011zty}
 Among many studies, see, for example, ATLAS Collaboration,
``Calibration of the ATLAS hadronic barrel calorimeter TileCal using 2008, 2009 and 2010 cosmic rays data,''
ATL-TILECAL-PUB-2011-001.


\bibitem{S_Banerjee}
Sunanda Banerjee (CMS), private communication.

\bibitem{Steerenberg}

``Status of the LHC" talk by R.~ Steerenberg  in the 12th LHCP conference, June, 2024. 

Available online at ~\href{https://indico.cern.ch/event/1253590/contributions/5814442/attachments/2869113/5022764/rs20240603_LHCP_LHC_Report.pdf}{https://indico.cern.ch/event/1253590/contributions/5814442/attachments/2869113/5022764/rs20240603-LHCP-LHC-Report.pdf}


\bibitem{Vannucci}
F.~Vannucci (private communication), as cited in~\cite{Petcov, Kopp:2007ai}

\bibitem{Petcov}
S. T. Petcov and T. Schwetz, Nucl. Phys. \textbf{B740}, 1 (2006).


\bibitem{Wen:2023ijf}
A.~Y.~Wen, C.~A.~Arg\"uelles, A.~Kheirandish and K.~Murase,
``Detecting High-Energy Neutrinos from Galactic Supernovae with ATLAS,''
Phys. Rev. Lett. \textbf{132}, no.6, 061001 (2024).

\bibitem{Ridky:2005mx}
J.~Ridky \textit{et al.} [DELPHI],
``Detection of muon bundles from cosmic ray showers by the DELPHI experiment,''
Nucl. Phys. B Proc. Suppl. \textbf{138}, 295-298 (2005)
doi:10.1016/j.nuclphysbps.2004.11.066

\bibitem{L3:2004sed}
P.~Achard \textit{et al.} [L3],
``Measurement of the atmospheric muon spectrum from 20-GeV to 3000-GeV,''
Phys. Lett. B \textbf{598}, 15-32 (2004)
doi:10.1016/j.physletb.2004.08.003
[arXiv:hep-ex/0408114 [hep-ex]].



\bibitem{ALICE:2015wfa}
J.~Adam \textit{et al.} [ALICE],
``Study of cosmic ray events with high muon multiplicity using the ALICE detector at the CERN Large Hadron Collider,''
JCAP \textbf{01}, 032 (2016)
doi:10.1088/1475-7516/2016/01/032
[arXiv:1507.07577 [astro-ph.HE]].


\bibitem{CMS:2010yju}
V.~Khachatryan \textit{et al.} [CMS],
``Measurement of the Charge Ratio of Atmospheric Muons with the CMS Detector,''
Phys. Lett. B \textbf{692}, 83-104 (2010)
doi:10.1016/j.physletb.2010.07.033
[arXiv:1005.5332 [hep-ex]].

\bibitem{ATLAS:2010apf}
G.~Aad \textit{et al.} [ATLAS],
``Studies of the performance of the ATLAS detector using cosmic-ray muons,''
Eur. Phys. J. C \textbf{71}, 1593 (2011)
doi:10.1140/epjc/s10052-011-1593-6
[arXiv:1011.6665 [physics.ins-det]].

\bibitem{ATLAS:2020ofx}
 ATLAS Collaboration,
``Identification of very-low transverse momentum muons in the ATLAS experiment,''
ATL-PHYS-PUB-2020-002.

\bibitem{ATLAS:1997ad}
 [ATLAS],
``ATLAS muon spectrometer: Technical design report,''
CERN-LHCC-97-22.

\bibitem{RPC_Table}
G.~Aad \textit{et al.} [ATLAS],
``The ATLAS experiment at the CERN Large Hadron Collider: a description of the detector configuration for Run~3,''
JINST \textbf{19}, no.05, P05063 (2024).

%
\bibitem{Formaggio:2012cpf}
J.~A.~Formaggio and G.~P.~Zeller,
``From eV to EeV: Neutrino Cross Sections Across Energy Scales,''
Rev. Mod. Phys. \textbf{84}, 1307-1341 (2012).



\bibitem{Thomson:2013zua}
M.~Thomson,
``Modern particle physics,''
Cambridge University Press, 2013.

\bibitem{ATLAS:2016lqx}
G.~Aad \textit{et al.} [ATLAS],
``Muon reconstruction performance of the ATLAS detector in proton\textendash{}proton collision data at $\sqrt{s}$ =13 TeV,''
Eur. Phys. J. C \textbf{76}, no.5, 292 (2016)
doi:10.1140/epjc/s10052-016-4120-y
[arXiv:1603.05598 [hep-ex]].

\bibitem{ATLAS:2020auj}
G.~Aad \textit{et al.} [ATLAS],
``Muon reconstruction and identification efficiency in ATLAS using the full Run 2 $pp$ collision data set at $\sqrt{s}=13$ TeV,''
Eur. Phys. J. C \textbf{81}, no.7, 578 (2021)
doi:10.1140/epjc/s10052-021-09233-2
[arXiv:2012.00578 [hep-ex]].


\bibitem{James:1980my}
F.~James and M.~Roos,
``Errors on Ratios of Small Numbers of Events,''
Nucl. Phys. B \textbf{172}, 475-480 (1980)

\bibitem{Helene:1983ph}
O.~Helene,
``Errors in Experiments With Small Number of Events,''
Nucl. Instrum. Meth. A \textbf{228}, 120 (1984)


\bibitem{Latitude}
We thank Ranjan Laha for raising this point, and M.~V.~N.~ Murthy for clarifying it during the talk by one of the authors, S.~Mukhopadhyay, at the TAPP 2024 conference at IMSc., Chennai.



\bibitem{Lohmann:1985qg}
W.~Lohmann, R.~Kopp and R.~Voss,
``Energy Loss of Muons in the Energy Range 1-{GeV} to 10000-{GeV},''
doi:10.5170/CERN-1985-003

\bibitem{Groom:2001kq}
D.~E.~Groom, N.~V.~Mokhov and S.~I.~Striganov,
``Muon stopping power and range tables 10-MeV to 100-TeV,''
Atom. Data Nucl. Data Tabl. \textbf{78}, 183-356 (2001);
Tables for muon energy loss are also available at
http://pdg.lbl.gov/2009/AtomicNuclearProperties/.


\bibitem{Erkoca:2009by}
A.~E.~Erkoca, M.~H.~Reno and I.~Sarcevic,
``Muon Fluxes From Dark Matter Annihilation,''
Phys. Rev. D \textbf{80}, 043514 (2009).

\bibitem{Covi:2009xn}
L.~Covi, M.~Grefe, A.~Ibarra and D.~Tran,
``Neutrino Signals from Dark Matter Decay,''
JCAP \textbf{04}, 017 (2010).

\bibitem{Feldstein:2010xe}
B.~Feldstein and A.~L.~Fitzpatrick,
``Discovering Asymmetric Dark Matter with Anti-Neutrinos,''
JCAP \textbf{09}, 005 (2010)

\bibitem{Fukuda:2014xqa}
H.~Fukuda, S.~Matsumoto and S.~Mukhopadhyay,
``Asymmetric dark matter in early Universe chemical equilibrium always leads to an antineutrino signal,''
Phys. Rev. D \textbf{92}, no.1, 013008 (2015)

\bibitem{Gandhi:2004bj}
R.~Gandhi, P.~Ghoshal, S.~Goswami, P.~Mehta and S.~U.~Sankar,
``Earth matter effects at very long baselines and the neutrino mass hierarchy,''
Phys. Rev. D \textbf{73}, 053001 (2006)
doi:10.1103/PhysRevD.73.053001
[arXiv:hep-ph/0411252 [hep-ph]].

\bibitem{Esteban:2024eli}
I.~Esteban, M.~C.~Gonzalez-Garcia, M.~Maltoni, I.~Martinez-Soler, J.~P.~Pinheiro and T.~Schwetz,
``NuFit-6.0: updated global analysis of three-flavor neutrino oscillations,''
JHEP \textbf{12}, 216 (2024)
doi:10.1007/JHEP12(2024)216
[arXiv:2410.05380 [hep-ph]].

\end{thebibliography}
\end{document}